\documentclass[
    ,final            
  ]
  {aipproc}

\layoutstyle{8x11double}

\usepackage{natbib}

\begin{document}

\title{A future very-high-energy view of our Galaxy}


\author{S. Funk}{
  address={Kavli Institute for Particle Astrophysics and Cosmology,
    Stanford, CA 94025, USA}
}
\author{J.A. Hinton}{
  address={School of Physics \& Astronomy, University of Leeds, Leeds,
    LS2 9JT, UK}
}
\author{G. Hermann}{
  address={Max-Planck Institut f{\"u}r Kernphysik, P.O. Box 103980,
    D-69029, Heidelberg, Germany}
}
\author{S. Digel}{
  address={Kavli Institute for Particle Astrophysics and Cosmology,
    Stanford, CA-94025, USA}
}

\begin{abstract}
  The survey of the inner Galaxy with H.E.S.S.~\citep{HESS:scanpaper1,
    HESS:scanpaper2} was remarkably successful in detecting a wide
  range of new very-high-energy gamma-ray sources. New TeV gamma-ray
  emitting source classes were established, although several of the
  sources remain unidentified, and progress has been made in
  understanding particle acceleration in astrophysical sources. In
  this work, we constructed a model of a population of such
  very-high-energy gamma-ray emitters and normalised the flux and size
  distribution of this population model to the H.E.S.S.-discovered
  sources. Extrapolating that population of objects to lower flux
  levels we investigate what a future array of imaging atmospheric
  telescopes (IACTs) such as AGIS or CTA might detect in a survey of
  the Inner Galaxy with an order of magnitude improvement in
  sensitivity. The sheer number of sources detected together with the
  improved resolving power will likely result in a huge improvement in
  our understanding of the populations of galactic gamma-ray sources.
  A deep survey of the inner Milky Way would also support studies of
  the interstellar diffuse gamma-ray emission in regions of high
  cosmic-ray density. In the final section of this paper we
  investigate the science potential for the Galactic Centre region for
  studying energy-dependent diffusion with such a future array.
\end{abstract}

\maketitle


\section{Building the population}

The first step in the simulation of a future survey of the Galactic
plane is to establish a model for the source population. For this
purpose, the sources detected in the H.E.S.S.\ survey of the inner
Galaxy ($\pm$30$^{\circ}$) were assumed to be attributable to a single
population and the following observables were compared between these
sources and a synthetic population:

\begin{itemize}
\item Number of detected sources
\item Galactic longitude and latitude distributions
\item Gamma-ray flux distribution
\item Angular size distribution
\end{itemize}

\begin{figure}
  \includegraphics[width=.4\textwidth]{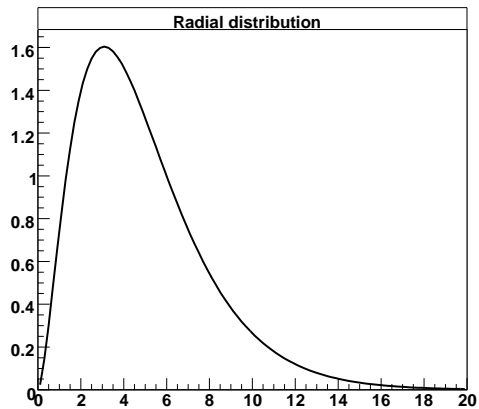}
  \caption{Radial distribution of sources in the SNR model in
    galactocentric coordinates. The units on the y-axis are
    arbitrary. }\label{fig::Fig1}
\end{figure}

Two different source population models have so far been tested
separately: an SNR-type and a PWN-type model. In the following we will
focus on the SNR population model.  To match the number of sources,
the frequency of SNe in our Galaxy was fixed to 10 per century. For
the distribution of Galactic longitudes and latitudes of the
underlying population we used a 3-dimensional distribution of sources
in our Galaxy. The radial distribution were taken from a model of of
SNRs in our Galaxy (see Fig.~\ref{fig::Fig1}) as given
in~\citep{CaseBhattacharya} and the height distribution above the
galactic disc was assumed to be exponential with a scale height of
30~kpc, approximately matched to the typical scale height of molecular
gas.

For the distribution of gamma-ray luminosities we used the model of
total explosion energies of Supernovae in our Galaxy as given
by~\citep{Sveshnikova} with the efficiency of converting that total
explosion energy into gamma rays as a single adjustable parameter. The
linear size of the model sources as a function of time was taken from
the Sedov-solution for SNRs~\citep{Sedov, Taylor}. Using a
distribution of densities of interstellar gas proportional to $n^{-1}$
around the SNRs, the gamma-ray flux from $\pi^0$-emission was then
calculated based on the prescription of Drury, Aharonian \&
V{\"o}lk~\citep{DAV1994}. As mentioned before, in this model, the
adjustable parameters (to match the H.E.S.S.\ distributions) are:

\begin{itemize}
\item SNR TeV gamma-ray lifetime
\item Efficiency of transferring explosion energy into kinetic energy
  of protons 
\end{itemize}

With these parameters a distribution of sources based on the SNR-model
can be constructed and compared to the above mentioned H.E.S.S.\
distributions. 

\section{Matching the H.E.S.S.\ population}
\begin{figure}
  \includegraphics[width=.48\textwidth]{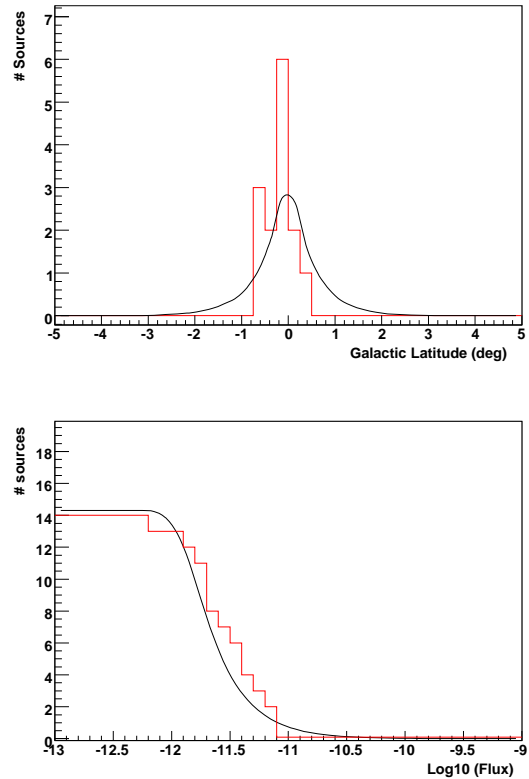}
  \caption{Top: Distribution of Galactic latitudes for model (black)
    and H.E.S.S.\ data (red). Bottom: Log(N)-Log(S) distributions for
    model (black) and H.E.S.S.\ data (red).}\label{fig::Fig2}
\end{figure}

The distributions of gamma-ray fluxes, galactic latitudes and angular
sizes of the H.E.S.S. sources in the inner Galaxy were fitted to the
respective distributions from the population. It seems clear from the
fit (as already suggested in the initial publication about the
H.E.S.S.\ survey~\citep{HESS:scanpaper2}), that the scale height of
the sources above the Galaxy must be rather small to match the narrow
distribution of H.E.S.S.\ sources in Galactic
latitude. Figures~\ref{fig::Fig2} show the distribution in Galactic
latitude and in Log($N$)-Log($S$) for the model and the H.E.S.S.\
data. The H.E.S.S. Log($N$)-Log($S$) distribution follows a power-law
with index $\sim$-1. A reasonable agreement between the model and the
H.E.S.S.\ data can be achieved with the following parameters:
\begin{itemize}
\item Number of SN explosions/century: 10
\item SNR TeV gamma-ray lifetime: 10$^4$ years
\item Scale height of the SNR distribution: 30 pc
\item Efficiency of converting explosion energy into cosmic rays: 9\%
\item Average density of surrounding medium: 0.5 cm$^{-3}$
\end{itemize}

With this population of sources we can reproduce the distribution of
sources and with this and the H.E.S.S.\ exposure map of the Inner
Galaxy and the H.E.S.S.\ background we can simulate the survey of the
inner Galaxy with both a H.E.S.S.-like and a future instrument.

\section{A future view of the inner Galaxy}
Using the population model derived in the previous section and
extrapolating this population to lower fluxes, we can now make an
educated guess at  what a future TeV gamma-ray instrument might be able to detect in a
survey of the inner Galaxy. Starting from the model population, we have to
include the instrument response function, i.e., the effective area and
the point-spread function of the instrument (as adjustable
parameters) to predict the number of detected gamma rays for the given
population model. We use the H.E.S.S. residual (photon-like) background
weighted by an adjustable parameter which denotes how much better the
background-rejection of a future instrument will be relative to the H.E.S.S.\
system. In addition to the population of sources we also add diffuse
gamma-ray emission from $\pi^0$-interactions using
CO maps~\citep{Dame:12CO} and assuming solar system cosmic ray (CR)
densities throughout the Galaxy.

\begin{figure}
  \includegraphics[width=.99\textwidth]{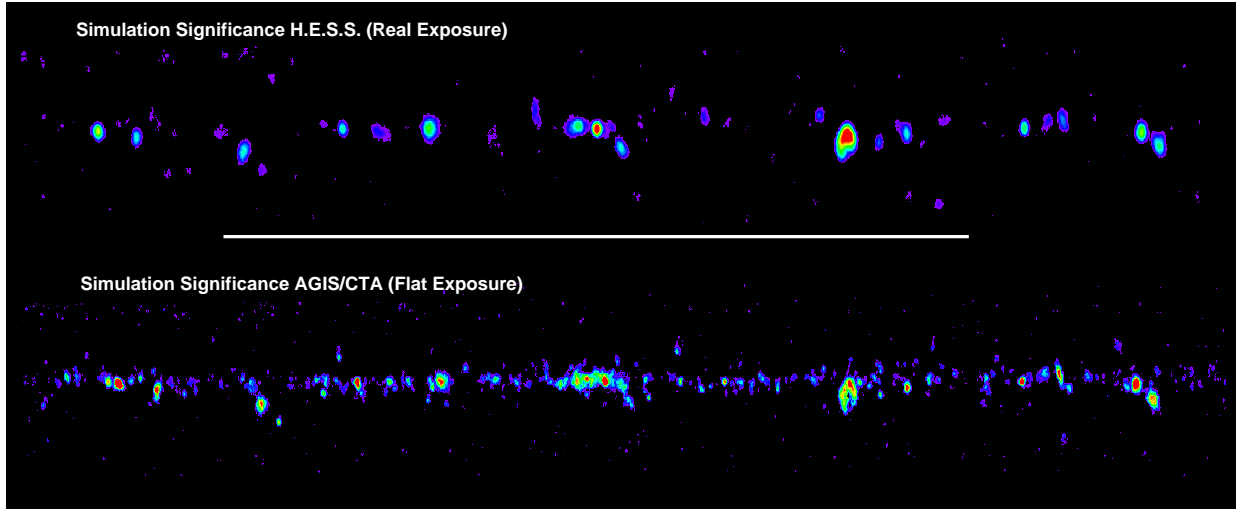}
  \caption{Significance Maps of the inner $l= \pm 30^{\circ}$ and $b =
    \pm 3^{\circ}$ for the population model described in this paper
    using the H.E.S.S.\ background. Top: for an array with the
    H.E.S.S. survey exposure and the H.E.S.S. angular resolution and
    background rejection. Bottom: for an array with a factor of 10
    larger area, a factor of 2 better background rejection and a
    factor of 2 improved angular resolution (resulting in a factor of
    ~9 improvement in sensitivity).}\label{fig::Fig3}
\end{figure}

Figure~\ref{fig::Fig3} shows simulated maps for the population of
sources derived in the previous section, one for an instrument with
H.E.S.S. effective areas, background rejection and angular resolution
(top) - it should be noted that in this case instead of using the
sources from the population, we used the measured parameters of the
H.E.S.S.\ sources in the inner Galaxy (in terms of flux normalisation,
spectral power-law index, position and extension) to enhance the
similarity to the well-known H.E.S.S.\ survey
picture~\citep{HESS:scanpaper2}.  This is why, e.g., we see a
shell-type SNR at the position of RX\,J1713.7-3946.  Nevertheless, the
distribution of these parameters for the population model matches the
distribution of H.E.S.S.\ sources as derived in the previous
section. The bottom plot of Figure~\ref{fig::Fig3} shows a simulated
significance map for a future survey of the inner Galaxy. Again, for
the brightest sources the parameters of the H.E.S.S. sources have been
used, whereas to extrapolate to lower fluxes we used the sources from
the population model. For this bottom plot, the exposure is a maximum
of 5 hours at any given point on the survey map was required (i.e., a
more even exposure than in the original H.E.S.S.\ survey). For this
particular map, the effective area of the future array was 10 times
that of H.E.S.S.\ and both the angular resolution and the background
rejection were conservatively improved only by a factor of 2,
resulting in an overall sensitivity increase by a factor of $2 \times
\sqrt 10 \times \sqrt 2 \sim 9$.

\section{Probing Energy-Dependent diffusion}

\begin{figure}
  \includegraphics[width=.45\textwidth]{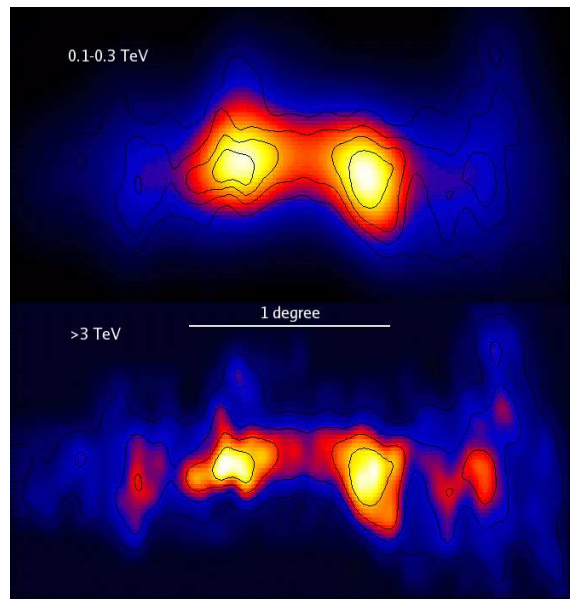}
  \caption{Simulated gamma-ray map in two energy bands in the Galactic
    Centre region, assuming energy-dependent diffusion of cosmic rays and a
    burst-like source in the center.}\label{fig::Fig4}
\end{figure}

In the plots shown in the previous paragraph we have integrated over
the whole energy range of a future Cherenkov array. A natural
extension to that scheme would be to split the energy range into
several bands to see how well spectral parameters can be measured. As
a first example of such an approach, we split the simulated diffuse
emission from interaction of cosmic rays within interstellar gas in
the Galactic Center region into several energy bands and show in
Figure~\ref{fig::Fig4} only the highest and the lowest range (0.1-0.3
TeV, and $E$ $>$ 3 TeV). With this approach we can test how well a future
Cherenkov instrument might study energy-dependent diffusion - assuming
a central source illuminates the clouds in the GC region.

For Figure~\ref{fig::Fig4}, CS data have been used to trace the target
material distribution (note that the unknown z-distribution was
assumed to be flat). The CR distribution was chosen as a
3-dimensional Gaussian with energy dependent width, corresponding to
burst-like injection in the past (e.g. a SN explosion). The diffusion
coefficient normalisation at 1 TeV was chosen to match the H.E.S.S.\
data in the GC region and an energy dependence of the form $D(E) \sim
E^{0.6}$ was assumed. Using a power-law injection energy spectrum of
the CRs we derived the spatial distribution of CRs in the
GC-region. Using this distribution of CRs we calculated the
distribution of $\pi^0$-decay gamma-rays following the parametrisation
of~\citet{Kelner:2006p218}. The gamma-ray maps shown in
Figure~\ref{fig::Fig4} have been smeared with the estimated
(energy-dependent) angular resolution of a CTA/AGIS-like instrument
and the probability density distribution for gamma-ray detection has
been plotted in the two energy bands specified above. A missing step
in this study (to be added in a future step) is the addition of
background and the sampling of photons with Poisson statistics.

\section{Summary}
This study provides a first glimpse of what a future array of
ground-based Cherenkov telescopes such as AGIS or CTA might be able to
study in the inner part of the Galaxy. We have constructed a
population model based on a distribution of SNR-like sources that is
able to explain the global properties of the H.E.S.S.-detected
sources. Extrapolating this population to gamma-ray flux values below
the H.E.S.S.-sensitivity gives a prediction of the richness of objects
that we might be able to study with a future instrument. Depending on
the exact parameters of the simulation we expect $\sim 300$ sources
above the flux sensitivity limit of CTA/AGIS for the inner
30$^{\circ}$ of the Galaxy. Next steps will include trade-off studies
between angular resolution, energy threshold and effective area to
guide the choice of array parameters. Detailed studies are needed for
a wide range of astrophysics topics to educate such a parameter
choice.  We have begun work on the specific area of the
energy-dependent diffusion in the Galactic Centre region, as a step
towards this.

\begin{theacknowledgments}
  The authors would like to thank the members of the H.E.S.S.\, CTA
  and AGIS collaborations for help and interesting discussions.
\end{theacknowledgments}



\bibliographystyle{aipproc}   

\bibliography{Gamma2008_SimulatedSurvey}

\IfFileExists{\jobname.bbl}{}
 {\typeout{}
  \typeout{******************************************}
  \typeout{** Please run "bibtex \jobname" to optain}
  \typeout{** the bibliography and then re-run LaTeX}
  \typeout{** twice to fix the references!}
  \typeout{******************************************}
  \typeout{}
 }

\end{document}